\documentclass{INTERSPEECH2023}
\usepackage{amsmath,graphicx}
\usepackage{multirow}
\usepackage{algorithm}
\usepackage[dvipsnames]{xcolor} % !!!!!!!!!!!!!!!!!!!!!!!!!!!!!!!!!!!
\usepackage[noend]{algpseudocode}
\usepackage{amsfonts}

\makeatletter

\def\algbackskip{\hskip-\ALG@thistlm}
\makeatother

\interspeechcameraready

\title{Defense Against Adversarial Attacks on Audio DeepFake Detection}
\name{Piotr Kawa$^{1*}$, Marcin Plata$^{1*}$, Piotr Syga$^1$ \thanks{* equal contribution}}
\address{
  $^1$Wroc{\l}aw University of Science and Technology, Wroc{\l}aw, Poland
}
\email{\{piotr.kawa, marcin.plata, piotr.syga\}@pwr.edu.pl}

\begin{document}

\maketitle
 
\begin{abstract}

Audio DeepFakes (DF) are artificially generated utterances created using deep learning, with the primary aim of fooling the listeners in a highly convincing manner.
Their quality is sufficient to pose a severe threat in terms of security and privacy, including the reliability of news or defamation.
Multiple neural network-based methods to detect generated speech have been proposed to prevent the threats.
In this work, we cover the topic of adversarial attacks, which decrease the performance of detectors by adding superficial (difficult to spot by a human) changes to input data. Our contribution contains evaluating the robustness of 3 detection architectures against adversarial attacks in two scenarios (white-box and using transferability) and enhancing it later by using adversarial training performed by our novel adaptive training. Moreover, one of the investigated architectures is RawNet3, which, to the best of our knowledge, we adapted for the first time to DeepFake detection.
\end{abstract}
\noindent\textbf{Index Terms}: audio DeepFakes, DeepFake detection, adversarial attacks, adversarial training

\section{Introduction}

The term DeepFakes (DF) describes a set of methods that allow the creation of artificial speech. They may rely on generating completely new sentences with Text-To-Speech or Voice-Cloning or transferring the properties of the victim's voice to the attacker's speech using Voice-Conversion~\cite{survey-df}. DF is a similar problem to voice spoofing~\cite{spoofing1,spoofing2}, yet significant differences exist between the two~\cite{asvspoof_2021_2}. Not only is the deceived subject different, but both problems differ in the types of attacks performed against the detection systems. One of the most typical spoofing attacks uses a previously recorded sample of a legitimate user voice (i.e., replay attack). Conversely, this is not an attack in the sense of DeepFakes. Moreover, multiple end-to-end attacks on multimodal features are used in DF, e.g., FakeAVCeleb dataset~\cite{fakeavceleb}.
The DF detection methods are focal --- the attacks may not only allow impersonation but also hinder the credibility of a medium that cites fake news or tarnish the victim's reputation if their voice has been used in a compromising recording. These methods can also afford unauthorized access to a system.

As a countermeasure, a set of audio DF detection methods that assess the validity of the utterances has been proposed. Currently, most solutions are based on deep neural networks~\cite{rawnet2,rawnet-3,lcnn,specrnet}. 
Existing architectures obtain good results in benchmarks; however, this may not reflect real-life scenarios due to the limited number and similarity of the datasets. 
For instance, the generalization of detectors to unseen generation methods is still unsatisfactory~\cite{attack-agnostic-dataset,does-df-generalize}. 
Defense against modified media results in the new legal regulations~\cite{eu_deepfakes} obliging content providers to implement self-regulatory procedures against disinformation such as DF.

In this paper, we lay our interest in another critical danger for DF detection systems --- adversarial attacks (AA), since the most efficient detection methods rely on deep learning approaches that are vulnerable to adversarial attacks (cf.~\cite{fgsm}). By adding small, often imperceptible, perturbations to input data, these attacks can cause the neural network to make an incorrect prediction. A limited investigation of the robustness against AA for audio spoofing detection was conducted~\cite{spoofingAdversarial,9053643} and in more detail for speaker and speech recognition~\cite{schonherr2018adversarial,cho22c_interspeech}. Moreover, the influence of the adversarial budget and transferability of adversarial training in the ASR environment have been investigated~\cite{AA-ASR3,AA-ASR4}. To the best of our knowledge, it has never been analyzed for audio DF detection.

We show that by using AAs, even with minimal knowledge of the adversary, we can conduct a successful attack on the DF detection system, significantly decreasing its performance.
Later, we provide measures to decrease this negative impact on the system. 
In Sect.~\ref{sec:white-box}, we measure the performance of white-box attacks using multiple AAs. Sect.~\ref{sec:transferability} concerns evaluating the transferability mechanism, which generates an attack using different models. 
Due to this, the adversary does not need access to the targeted model --- it allows us to evaluate the performance of an attack scenario with limited knowledge of the adversary (resembling a black-box attack). Additionally, we limit the joint preprocessing steps between the attack and targeted models to only common operations; hence the adversary does not need to know the targeted preprocessing pipeline.
In Sect.~\ref{sec:adversarial-training}, we increase the robustness of the classifiers against AA using our novel adaptive adversarial training and show its effectiveness in the ablation study.
Our codebase and samples related to this research are available on GitHub: \url{github.com/piotrkawa/audio-deepfake-adversarial-attacks}.

\section{Detection models and dataset}\label{sec:format}

In our work, we consider 3 audio DeepFake detection models -- LCNN~\cite{lcnn}, RawNet3~\cite{rawnet-3}, and SpecRNet~\cite{specrnet}.
Our motivation is the following --- LCNN is one of the most popular DF detectors, RawNet3 is a follow-up of RawNet2~\cite{rawnet2}, an architecture providing high results in many speech classification and DF detection benchmarks, whereas SpecRNet provides appropriate results using minimal resources~\cite{df-captcha}.
Please note that, to the best of our knowledge, we are the first to adapt RawNet3 (originally used in speaker recognition) to the DF detection problem.
The sizes of the detectors differ, containing 467,425 (LCNN), 15,496,197 (RawNet3), and 277,963 (SpecRNet) trainable parameters. Treating SpecRNet as a baseline, LCNN requires 52\%, and RawNet3 --- 608\%,  more time to process. RawNet3, with a batch size of 32, requires 11,766~MB of GPU VRAM, whereas LCNN --- 2,300~MB and SpecRNet --- 1,299~MB. 

We use data composed of 3 audio DF datasets: ASVspoof2021 (DF subset)~\cite{asvspoof_2021}, FakeAVCeleb~\cite{fakeavceleb} and WaveFake~\cite{wavefake}. This contributed to 41,217 real and 702,269 fake samples generated using over 100 DF methods. The reason behind combining multiple datasets is that recent works have shown that the DF solutions do not generalize well on attacks~\cite{attack-agnostic-dataset,does-df-generalize}. Thus one of the techniques to prevent this issue is to combine multiple datasets as presented. 
Our dataset is based on a random selection of 50,000 fake samples for training and 5,000 fake samples each for validation and test subsets. This results in around 500 training and 50 validation and test samples per DF generation method. Real samples are spread across particular subsets with the same proportion (10:1:1), and then the training subset is oversampled to eliminate the disparity of classes. The quantity of the data surpasses recent works~\cite{cho22c_interspeech}.

The audio samples are preprocessed by resampling them to 16~kHz mono, removing silences longer than 0.2~s, and then either trimming or padding (by repeating) to about 4~s (which is a common preprocessing technique~\cite{wavefake, rawnet2,attack-agnostic-dataset,specrnet}).

Training detection models lasted 10 epochs, each ending with validation --- the checkpoint with the highest accuracy was selected for the test procedure. Each process used the same randomness seed. We trained using a binary cross-entropy loss function and a batch size of 128. Learning rates were equal to $10^{-4}$ for LCNN and SpecRNet and $10^{-3}$ for RawNet3. SpecRNet used a weight decay of $10^{-4}$ whereas RawNet3 of $5\cdot 10^{-5}$. Additionally, RawNet3 was trained using SGDR scheduling procedure~\cite{sgdr} restarting after every epoch. LCNN and SpecRNet used a linear frequency cepstral coefficients (LFCC) front-end~\cite{lfcc_mfcc} as an input layer with 25~ms Hann window, 10~ms window shift, 512 FFT points and 80 coefficients~\cite{fakeavceleb}, whereas RawNet3 analyzed raw audio. We reported test results using an Equal Error Rate (EER), typically used in tasks like biometric verification or DF detection. We based our codebase on~\cite{attack-agnostic-dataset}.

\section{White-box attack benchmark}\label{sec:white-box}

We evaluated the robustness of DF detection models against adversarial attacks, including Fast Gradient Signed Method (FGSM)~\cite{fgsm}, Projected Gradient Descent (PGD, $L_2$ variant)~\cite{pgdl2}, and Fast Adaptive Boundary (FAB) attack~\cite{fab}. 
These methods are one of the most popular adversarial attacks, extensively studied in the literature (including audio processing and computer vision), and represent various approaches to the problem~\cite{Hendrycks_2021_ICCV,NEURIPS2021_e19347e1,Liu_2022_CVPR}.
FGSM and FAB were based on a $L_{\infty}$ norm, whereas PGDL2 was on $L_2$. 
We used 3 variants of each attack differing in the parameters. 
As DeepFakes aim to deceive people, the attacks on the detectors should comply with two seemingly mutually exclusive properties --- the samples should have as least distortions as possible to fool humans while having enough distortions to fool the detection system (attacks with less impact tend to fail more often). 
PGDL2 used 10 steps and $\varepsilon\in\{0.1, 0.15, 0.20\}$, FGSM $\varepsilon\in \{0.0005, 0.00075, 0.001\}$, whereas FAB utilized $\eta\in\{10,20,30\}$.
These parameters were selected based on human evaluation. Having the quality scores of samples attacked with various parameters given by 10 volunteers, we selected the attack levels at which distortions were imperceptible.
We added two parameters with increasing levels of manipulation, resulting in a noticeable yet acceptable level of distortion, still allowing to understand an utterance and identify a speaker.
In samples generated with higher values of $\eta$ and $\varepsilon$ parameters, slight background noise could be heard. We additionally report mean mel-cepstral distortion~\cite{mcd} of successfully attacked samples (ordered by increasing distortion levels) --- 2.602, 3.460, 4.182 for FGSM attack, then 2.445, 3.637, 4.500 for PGDL2 attack and 2.311, 3.508, 4.469 for FAB attack.

White-box adversarial attacks assume an adversary has access to the targeted model, including its weights.
Our benchmark started with evaluating the networks on the original (non-attack) data. Then, we evaluated each of them on the same full test subset altered using AAs. Attacks were performed directly on the audio files.

   \begin{table}[h]
    \centering
    \caption{Comparison of EERs on the samples altered by the white-box adversarial attack. The values in bold denote the greatest EERs for each attack.}
    \label{tab:standard_and_white_box}     
    \begin{tabular}{c c c c}
    \hline
        \multicolumn{4}{c}{\textbf{FGSM}} \\
        \hline
        Model &$\varepsilon=0.0005$&$\varepsilon=0.00075$&$\varepsilon=0.001$\\
        % \hline
        LCNN & 0.8126 & 0.7686 & 0.7072 \\ 
        SpecRNet & \textbf{0.9676} & \textbf{0.9386} & \textbf{0.8716} \\ 
        RawNet3 &  0.2118 & 0.2095 & 0.2112\\ 
    \hline
    \multicolumn{4}{c}{\textbf{PGDL2}} \\
        \hline
        Model &$\varepsilon=0.1$ &   $\varepsilon=0.15$& $\varepsilon=0.20$\\
        % \hline
        LCNN & 0.9501 & 0.9716&  0.9791\\ 
        SpecRNet & \textbf{0.9803} & \textbf{0.9865} & \textbf{0.9905} \\ 
        RawNet3 & 0.3102 & 0.4567  &  0.5499 \\ 
    \hline
    \multicolumn{4}{c}{\textbf{FAB}} \\
        \hline
        Model &$\eta=10$ &   $\eta=20$& $\eta=30$\\
        % \hline
        LCNN & 0.6305 & 0.6315&   0.6321\\
        SpecRNet &0.6833 & 0.7394& 0.7614 \\
        RawNet3 &  \textbf{0.7430}   & \textbf{0.8265}   & \textbf{0.8549} \\
        \hline
     \end{tabular}   
    \end{table}

All models achieve high performance in the non-attack setting --- EERs of 0.0227, 0.0258, and 0.0180 for, respectively, LCNN, SpecRNet, and RawNet3. 
Table~\ref{tab:standard_and_white_box} shows that the AAs significantly decrease models' performance making them ineffectual. 
The performances differ depending on the targeted model. The greatest impact is visible for SpecRNet, whereas RawNet3 is the most robust. There is no superior attack --- while LCNN and SpecRNet obtain the worst results with PGDL2 ($\varepsilon$ = 0.20),
the worst performance of RawNet3 is caused by FAB ($\eta$ = 30) simultaneously being the least effective for the rest. Interestingly, while PGDL2 and FAB get more destructive with higher parameters, FGSM does not. In fact, its effectiveness decreases for LCNN and SpecRNet. The adversary should carefully select the method and its parameters to conduct the most damaging attack. The highest parameters do not necessarily guarantee the worst score of EER.

\section{Transferability benchmark}\label{sec:transferability}

Transferability scenario covers AA using models differing from the targeted ones --- without access to the targeted model or its weights. The field of audio DF detection is still relatively new, and there are few available datasets and architectures. Some networks, such as LCNN, are widely used, letting an adversary assume it could be used in the system. 
Moreover, the preprocessing used during training is well established and usually includes only resampling and duration standardization. Thus, using similar methods, the adversary could effectively attack a sample. However, if the exact parameters are not given, the adversary knows only the preprocessing operations and the datasets. Due to the limited information, this approach can be compared to the black-box setting.

The pipeline of transferability attacks differed from the previous one only in the used models, which are different for the training and the attack. 
Table~\ref{tab:transferability-eval} shows that the transferability attacks are less efficient than white-box; however, they still can significantly decrease the performances. We can attack LCNN using SpecRNet and vice versa, scoring average EERs of 0.2001 and 0.3236, which indicates adversarial attacks still have noticeable strength.
For both targeting RawNet3 and using it to attack, we observe low EERs, which means the model is robust against adversarial attacks and, at the same time, is inefficient as an attack model. 
We also investigate the transferability between LCNN and SpecRNet. Our first assumption was that it was related to their front-end (they both use LFCC). Due to their different structure, this is their main common component. However, results of attacks with mel-frequency cepstral coefficients (MFCC) front-end~\cite{lfcc_mfcc} achieve comparable results --- we conclude that the reason behind good transferability is the similarity in networks' sizes. 
Models, depending on their sizes, capture different patterns. Attack with a model of a different size does not necessarily modify artifacts crucial for the detection model.
   \begin{table}[h]
    \centering
    \caption{Transferability attacks results. TM denotes the target model, whereas AM --- attack model. We report parameter values in rows (in increasing order). The values in bold denote EERs between smaller architectures.}
    \label{tab:transferability-eval} 
    \begin{tabular}{c c c c c}
    \hline
        \multicolumn{5}{c}{FGSM} \\
        \hline
         TM & AM & $\varepsilon_{1}$ &  $\varepsilon_{2}$ & $\varepsilon_{3}$\\
 
        LCNN & SpecRNet &  \textbf{0.2500} & \textbf{0.2560} & \textbf{0.2530} \\ 
        LCNN & RawNet3 &  0.0680   & 0.0984 & 0.1352 \\ 
        
        SpecRNet & RawNet3 & 0.1039 & 0.1683 & 0.2061   \\ 
        SpecRNet & LCNN & \textbf{0.2542} & \textbf{0.2835} & \textbf{0.2993}  \\ 
        
        RawNet3 & LCNN &  0.0795 & 0.0948 & 0.1082   \\ 
        RawNet3 & SpecRNet &  0.0750 & 0.0901 & 0.1003  \\ \hline
        
        \multicolumn{5}{c}{PGDL2} \\
        \hline
        TM & AM &$\varepsilon_{1}$ &   $\varepsilon_{2}$& $\varepsilon_{3}$\\
        
        LCNN & SpecRNet &  \textbf{0.1625} & \textbf{0.2375} & \textbf{0.2908} \\ 
        LCNN & RawNet3 &  0.0322 & 0.0444 &	0.0630   \\ 
        
        SpecRNet & RawNet3 &  0.0746 & 0.1267 & 0.1916  \\
        SpecRNet & LCNN &  \textbf{0.3526} & \textbf{0.4429} & \textbf{0.4867} \\ 
        
        RawNet3 & LCNN &  0.0368 & 0.0813 & 0.1124   \\ 
        RawNet3 & SpecRNet &  0.0294 & 0.0549 & 0.0839  \\ \hline

        \multicolumn{5}{c}{FAB} \\
        \hline
        TM & AM &$\eta_{1}$ &   $\eta_{2}$& $\eta_{3}$\\
        
        LCNN & SpecRNet & \textbf{0.0591} & \textbf{0.1207} & \textbf{0.1710}  \\ 
        LCNN & RawNet3 & 0.0228	& 0.0228 & 	0.0228   \\ 
        
        SpecRNet & RawNet3 &  0.0249 & 0.0249 & 0.0249  \\ 
        SpecRNet & LCNN &  \textbf{0.2129} & \textbf{0.2778} & \textbf{0.3025} \\ 
        
        RawNet3 & LCNN &  0.1010 & 0.1149 & 0.1270   \\ 
        RawNet3 & SpecRNet &  0.0257 & 0.0408 & 0.0525  \\ \hline
     \end{tabular}
    \end{table}

\section{Adversarial Training}\label{sec:adversarial-training}

One of the most popular strategies to circumvent adversarial attacks is adversarial training~\cite{SzegedyZSBEGF13}. It relies on generating adversarial samples during the training and using them to train the neural network. It is considered one of the most effective techniques to increase robustness~\cite{ijcai2021p591}. We decided to conduct experiments with adversarial training for LCNN architecture, as it is one of two small models achieving weak robustness results against adversarial attacks. Moreover, LCNN is one of the most widely used algorithms in DF audio detection and spoofing tasks making it a good choice as this model can potentially be used in real-life detection systems. 

\subsection{Training Method}
\label{sec:training}

Default adversarial training procedure aims at increasing robustness against one particular attack. In most known scenarios, the model is fine-tuned by providing only adversarially attacked samples. 
Other scenarios rely on building a batch containing a mix of attacks and unchanged samples or using an ensemble of different attacks' parameters and models~\cite{tramer2018ensemble}.
These approaches do not consider the complexity of individual attacks, e.g., some may be simpler to detect due to background noise and thus easier to learn. To address the described problem, as well as other common problems, such as generalization and (catastrophic) overfitting~\cite{ijcai2021p591}, we propose a new variant of \textit{adaptive} training approach, in which we continuously analyze which attack is difficult to detect. To do this, we choose an attack (or skip it), making a decision based on a sampling vector $w \in \mathbb{R_{+}}^{N+1}$, that $\sum_{i=0}^{N} w_i = 1$, where $N$ is the number of attacks, and $w_0$ refers to a non-attacking scenario. 
We directly determine the effectiveness of the detector based on a loss result from a previous iteration. In Algorithm~\ref{pseudcode}, the pseudocode of the \textit{adaptive} approach is shown.

    \begin{algorithm}
    \caption{Adaptive update of the sampling weights vector.}\label{pseudcode}
    \hspace*{\algorithmicindent} \textbf{Input:} \\ 
    \hspace*{\algorithmicindent} $w$ --- $N + 1$-dimensional sampling vector, \\  
    \hspace*{\algorithmicindent} $loss_i$ --- loss result, \\ 
    \hspace*{\algorithmicindent} $i \in \{0, \dots, N\}$ --- (non-)attack index  \\
    \hspace*{\algorithmicindent} $c$ --- clip value (default: $1$),  \\
    \hspace*{\algorithmicindent}  $m$ --- momentum ($\frac{1}{5}$), \\
    \hspace*{\algorithmicindent} $p$ --- non-attack proportion ($\frac{1}{3})$. \\
    \hspace*{\algorithmicindent} \textbf{Output} updated sampling vector $w$  
    \begin{algorithmic}[1]
    \Procedure{AdaptiveUpdate}{}
    \State $w_i \gets m \cdot min(loss_i, c) + (1-m) \cdot w_i$    
    \State $s_w \gets \sum_{k=0}^{N} w_k$ 
    \State \For{$k \gets 0$ to $N$}
          \If {$k = 0$} 
            \State $r \gets p$
          \Else
            \State $r \gets \frac{1-p}{N}$
          \EndIf
          \State $w_k \gets \frac{1}{2} \cdot \frac{w_k}{s_w} + \frac{1}{2} \cdot r$
    \EndFor
    \State \Return $w$
    \EndProcedure
    \end{algorithmic}
    \end{algorithm}
    
We apply two control mechanisms to prevent an element $w_i$ of the sampling vector from being $w_i \gg w_j$ for $j \in \{0, \dots, N\} \setminus \{i\}$ resulting in choosing only one (non-)attack w.h.p.
First, we constrain the updated value $w_i$ with the clip value $c$, and additionally, we average the vector $w$ with a constant ratio $r$. Moreover, we use a constant parameter $p$ to define a proportion for sampling original samples since we aim to keep the model robust in the non-AA scenario. 
The training covered fine-tuning the baseline model for 10 epochs with the same training hyper-parameters (Sect.~\ref{sec:white-box}). The reason behind this approach was a problem with convergence when training from the start (a similar approach was used in \cite{cho22c_interspeech}). To avoid catastrophic overfitting, after every epoch, we validate the performance of the model using the formula: 
 \begin{equation}
 \label{eq:f1}
     e^{*} = \operatorname*{argmax}_{e\in \{1, \dots, 10\}} \frac{(N+1) \prod_{i=0}^{N}a^e_i}{\sum_{i=0}^{N}a^e_i},
 \end{equation}
where $a^e_i$ is an accuracy of the $i$-th (non-)attack index after $e$-th epoch. We choose the model after the epoch~$e^{*}$.

\subsection{Results}

The LCNN model fine-tuned using our adaptive adversarial training significantly improves the white-box scenario and increases the robustness for many attacks in the transferability benchmark. At the same time, its EER score calculated on original samples is equal to 0.0882, which is still satisfactory. 
This shows that robustness against AAs is possible without a significant decrease in performance for a standard scenario.
The results are presented in Table~\ref{tab:wb-at} and Table~\ref{tab:bb-at}. 

Adaptive adversarial training significantly increases the white-box scenario's robustness. The average EER decreased from 0.7870 to 0.1247. Regarding transferability, the average EER score also improved --- achieving 0.1133. The improvement shows primarily in the case of the attack with SpecRNet, as RawNet3 does not incur a substantial threat increase in the transferability scenario (Sect.~\ref{sec:transferability}).

   \begin{table}[h]
    \centering
    \caption{Adversarial training enhances the robustness against white-box AAs in case of all results (cf.~Table~\ref{tab:standard_and_white_box}).}
    \label{tab:wb-at}
    \begin{tabular}{c c c c}
    \hline
        \multicolumn{4}{c}{\textbf{FGSM}} \\
        \hline
        Model & $\varepsilon=0.0005$ &  $\varepsilon=0.00075$ & $\varepsilon=0.001$\\
        % \hline
        LCNN & \textbf{0.0985} & \textbf{0.1079} & \textbf{0.1144} \\ \hline

    \multicolumn{4}{c}{\textbf{PGDL2}} \\
        \hline
        Model &$\varepsilon=0.1$ &   $\varepsilon=0.15$& $\varepsilon=0.20$\\
        % \hline
        LCNN & \textbf{0.1245} & \textbf{0.1511} & \textbf{0.1870} \\ \hline

    \multicolumn{4}{c}{\textbf{FAB}} \\
        \hline
        Model &$\eta=10$ &   $\eta=20$& $\eta=30$\\
        % \hline
        LCNN & \textbf{0.0982} & \textbf{0.1116} & \textbf{0.1246} \\ \hline

     \end{tabular}
    \end{table}

   \begin{table}[h]
    \centering
    \caption{Transferability attacks results after adversarial training. TM denotes target model, AM --- attack model. Bold cells refer to results with enhanced EER (cf.~Table~\ref{tab:transferability-eval}).}
    \label{tab:bb-at}    
    \begin{tabular}{c c c c c}
    \hline
        \multicolumn{5}{c}{\textbf{FGSM}} \\
        \hline
        TM & AM & $\varepsilon_{1}$ &  $\varepsilon_{2}$ & $\varepsilon_{3}$\\
        % \hline
        LCNN & SpecRNet &  \textbf{0.1472} & \textbf{0.1515} & \textbf{0.1560} \\
        LCNN & RawNet3   & 0.0994  & 0.1064 & \textbf{0.1091} \\
        \hline
        \multicolumn{5}{c}{\textbf{PGDL2}} \\
        \hline
        TM & AM &$\varepsilon_{1}$ &   $\varepsilon_{2}$& $\varepsilon_{3}$\\
        % \hline
        LCNN & SpecRNet & \textbf{0.1162}  & \textbf{0.1247} &  \textbf{0.1434} \\
        LCNN & RawNet3  & 0.0876  & 0.0920 & 0.0945 \\

        \hline
        \multicolumn{5}{c}{\textbf{FAB}} \\
        \hline
        TM & AM &$\eta_{1}$ & $\eta_{2}$& $\eta_{3}$\\
        LCNN & SpecRNet & 0.1018 & \textbf{0.1158} & \textbf{0.1229} \\
        LCNN & RawNet3 &  0.0901 & 0.0902 & 0.0904 \\
        \hline
        
     \end{tabular}    
    \end{table}

\subsection{Ablation study}

\begin{figure}
    \centering
    \includegraphics[width=0.5\textwidth]{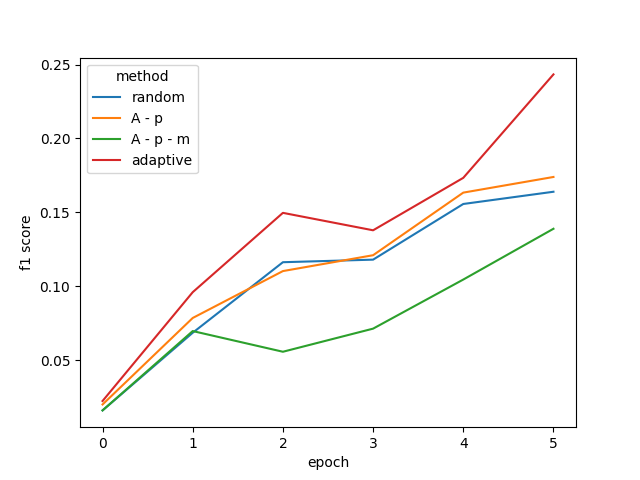}
    \caption{The comparison of F1 score for 4 training methods considered in the ablation study. Our adaptive method provides the best effectiveness.}
    \label{fig:abl}
\end{figure}

In order to demonstrate the effectiveness of our adaptive training,  we provide an analysis of four training cases: random, adaptive ($A$), adaptive without non-attack proportion $p$ ($A - p$), and simple adaptive without non-attack proportion $p$ and momentum $m$ ($A-p-m$). The last two are simplified cases of the adaptive method described in Sect.~\ref{sec:training}. The method $A-p$ differs from the Algorithm~\ref{pseudcode} in lines 6-10 as the sampling vector is updated by $w_k \gets \frac{w_k}{s_w}$. The method $A-p-m$ removes momentum from line 1 (it is equivalent to $m=1$). Note that we always apply value clipping, as otherwise, the sampling vector converges to a case with one dominant element that is chosen most frequently. 
We aim to achieve general robustness against all attacks and non-attack cases; thus, we propose an evaluation metric that is the multi-class F1 score calculated on the (non-)attack accuracies (similar to the formula~\ref{eq:f1}). The metric promotes cases where the achieved accuracies have a small variance. The results are presented in Figure~\ref{fig:abl}. The $A-p-m$ method achieves worse results than the random. One of the reasons is that in the initial training phase, the loss value is relatively high, causing the sampling vector to be unbalanced quickly (far from uniform), and only one or a few attacks are sampled w.h.p. To prevent this, two mechanisms have to be applied -- clipping and momentum. This leads to the results being only slightly better than the random. Finally, adding the non-attack proportion (i.e. using our adaptive version) clearly improves the F1 scores over the other cases achieving the overall best results. This ensures the model does not lose patterns to distinguish between real and fake samples.

\section{Conclusions}

In this work, we show that adversarial attacks severely threaten audio DeepFake detection methods. We present two scenarios of the attacks --- the white box and using the transferability mechanism. The second allows us to attack a model even with limited knowledge of the adversary, i.e., without information about the target model. Finally, we present a novel method of conducting adversarial training that generalizes well and is resistant to overfitting. This allows us to increase the robustness against many combinations of attacks in both white-box and transferability scenarios leading to acceptable results even after adversarial attacks. 

\section{Acknowledgments}
This work has been partially funded by Department of Artificial Intelligence, Wroc{\l}aw University of Science and Technology.

\bibliographystyle{IEEEtran}
\bibliography{mybib}

\begin{thebibliography}{10}
\providecommand{\url}[1]{#1}
\csname url@samestyle\endcsname
\providecommand{\newblock}{\relax}
\providecommand{\bibinfo}[2]{#2}
\providecommand{\BIBentrySTDinterwordspacing}{\spaceskip=0pt\relax}
\providecommand{\BIBentryALTinterwordstretchfactor}{4}
\providecommand{\BIBentryALTinterwordspacing}{\spaceskip=\fontdimen2\font plus
\BIBentryALTinterwordstretchfactor\fontdimen3\font minus
  \fontdimen4\font\relax}
\providecommand{\BIBforeignlanguage}[2]{{%
\expandafter\ifx\csname l@#1\endcsname\relax
\typeout{** WARNING: IEEEtran.bst: No hyphenation pattern has been}%
\typeout{** loaded for the language `#1'. Using the pattern for}%
\typeout{** the default language instead.}%
\else
\language=\csname l@#1\endcsname
\fi
#2}}
\providecommand{\BIBdecl}{\relax}
\BIBdecl

\bibitem{survey-df}
M.~Masood, M.~Nawaz, K.~M. Malik, A.~Javed, A.~Irtaza, and H.~Malik,
  ``Deepfakes generation and detection: State-of-the-art, open challenges,
  countermeasures, and way forward,'' \emph{Applied Intelligence}, pp. 1--53,
  2022.

\bibitem{spoofing1}
T.~Kinnunen, H.~Delgado, N.~Evans, K.~A. Lee, V.~Vestman, A.~Nautsch,
  M.~Todisco, X.~Wang, M.~Sahidullah, J.~Yamagishi, and D.~A. Reynolds,
  ``Tandem assessment of spoofing countermeasures and automatic speaker
  verification: Fundamentals,'' \emph{IEEE/ACM Transactions on Audio, Speech,
  and Language Processing}, vol.~28, pp. 2195--2210, 2020.

\bibitem{spoofing2}
X.~Wang and J.~Yamagishi, ``{Investigating Self-Supervised Front Ends for
  Speech Spoofing Countermeasures},'' in \emph{Proc. The Speaker and Language
  Recognition Workshop (Odyssey 2022)}, 2022, pp. 100--106.

\bibitem{asvspoof_2021_2}
H.~Delgado, N.~Evans, T.~Kinnunen, K.~A. Lee, X.~Liu, A.~Nautsch, J.~Patino,
  M.~Sahidullah, M.~Todisco, X.~Wang \emph{et~al.}, ``Asvspoof 2021: Automatic
  speaker verification spoofing and countermeasures challenge evaluation
  plan,'' \emph{arXiv preprint arXiv:2109.00535}, 2021.

\bibitem{fakeavceleb}
H.~Khalid, S.~Tariq, M.~Kim, and S.~S. Woo, ``Fake{AVC}eleb: A novel
  audio-video multimodal deepfake dataset,'' in \emph{Thirty-fifth Conference
  on Neural Information Processing Systems Datasets and Benchmarks Track (Round
  2)}, 2021.

\bibitem{rawnet2}
\BIBentryALTinterwordspacing
H.~Tak, J.~Patino, M.~Todisco, A.~Nautsch, N.~Evans, and A.~Larcher,
  ``End-to-end anti-spoofing with rawnet2,'' \emph{2021 IEEE International
  Conference on Acoustics, Speech and Signal Processing (ICASSP)}, Jun 2021.
  [Online]. Available: \url{http://dx.doi.org/10.1109/ICASSP39728.2021.9414234}
\BIBentrySTDinterwordspacing

\bibitem{rawnet-3}
J.~weon Jung, Y.~Kim, H.-S. Heo, B.-J. Lee, Y.~Kwon, and J.~S. Chung,
  ``{Pushing the limits of raw waveform speaker recognition},'' in \emph{Proc.
  Interspeech 2022}, 2022, pp. 2228--2232.

\bibitem{lcnn}
X.~Wu, R.~He, Z.~Sun, and T.~Tan, ``A light cnn for deep face representation
  with noisy labels,'' \emph{IEEE Transactions on Information Forensics and
  Security}, vol.~13, no.~11, pp. 2884--2896, 2018.

\bibitem{specrnet}
P.~Kawa, M.~Plata, and P.~Syga, ``Specrnet: Towards faster and more accessible
  audio deepfake detection,'' in \emph{2022 IEEE 21st International Conference
  on Trust, Security and Privacy in Computing and Communications
  (TrustCom)}.\hskip 1em plus 0.5em minus 0.4em\relax IEEE, 2022.

\bibitem{attack-agnostic-dataset}
------, ``{Attack Agnostic Dataset: Towards Generalization and Stabilization of
  Audio DeepFake Detection},'' in \emph{Proc. Interspeech 2022}, 2022, pp.
  4023--4027.

\bibitem{does-df-generalize}
N.~Müller, P.~Czempin, F.~Diekmann, A.~Froghyar, and K.~Böttinger, ``{Does
  Audio Deepfake Detection Generalize?}'' in \emph{Proc. Interspeech 2022},
  2022, pp. 2783--2787.

\bibitem{eu_deepfakes}
\BIBentryALTinterwordspacing
{European Commission}, ``{2022 Strengthened Code of Practice on
  Disinformation},'' 2022. [Online]. Available:
  \url{https://digital-strategy.ec.europa.eu/en/library/2022-strengthened-code-practice-disinformation}
\BIBentrySTDinterwordspacing

\bibitem{fgsm}
\BIBentryALTinterwordspacing
I.~J. Goodfellow, J.~Shlens, and C.~Szegedy, ``Explaining and harnessing
  adversarial examples,'' in \emph{3rd International Conference on Learning
  Representations, {ICLR}}, 2015. [Online]. Available:
  \url{http://arxiv.org/abs/1412.6572}
\BIBentrySTDinterwordspacing

\bibitem{spoofingAdversarial}
S.~Liu, H.~Wu, H.-Y. Lee, and H.~Meng, ``Adversarial attacks on spoofing
  countermeasures of automatic speaker verification,'' in \emph{2019 IEEE
  Automatic Speech Recognition and Understanding Workshop (ASRU)}, 2019, pp.
  312--319.

\bibitem{9053643}
H.~Wu, S.~Liu, H.~Meng, and H.-y. Lee, ``Defense against adversarial attacks on
  spoofing countermeasures of asv,'' in \emph{2020 IEEE International
  Conference on Acoustics, Speech and Signal Processing (ICASSP)}, 2020, pp.
  6564--6568.

\bibitem{schonherr2018adversarial}
L.~Sch{\"{o}}nherr, K.~Kohls, S.~Zeiler, T.~Holz, and D.~Kolossa, ``Adversarial
  attacks against automatic speech recognition systems via psychoacoustic
  hiding,'' in \emph{26th Annual Network and Distributed System Security
  Symposium, {NDSS} 2019, San Diego, California, USA, February 24-27,
  2019}.\hskip 1em plus 0.5em minus 0.4em\relax The Internet Society, 2019.

\bibitem{cho22c_interspeech}
S.~Joshi, S.~Kataria, Y.~Shao, P.~Żelasko, J.~Villalba, S.~Khudanpur, and
  N.~Dehak, ``{Defense against Adversarial Attacks on Hybrid Speech Recognition
  System using Adversarial Fine-tuning with Denoiser},'' in \emph{Proc.
  Interspeech 2022}, 2022, pp. 5035--5039.

\bibitem{AA-ASR3}
Y.~Shao, J.~Villalba, S.~Joshi, S.~Kataria, S.~Khudanpur, and N.~Dehak,
  ``{Chunking Defense for Adversarial Attacks on ASR},'' in \emph{Proc.
  Interspeech 2022}, 2022, pp. 5045--5049.

\bibitem{AA-ASR4}
H.~Tan, J.~Zhang, H.~Zhang, L.~Wang, Y.~Qian, and Z.~Gu, ``{NRI-FGSM: An
  Efficient Transferable Adversarial Attack for Speaker Recognition Systems},''
  in \emph{Proc. Interspeech 2022}, 2022, pp. 4386--4390.

\bibitem{df-captcha}
L.~Yasur, G.~Frankovits, F.~M. Grabovski, and Y.~Mirsky, ``Deepfake captcha: A
  method for preventing fake calls,'' \emph{arXiv preprint arXiv:2301.03064},
  2023.

\bibitem{asvspoof_2021}
J.~Yamagishi, X.~Wang, M.~Todisco, M.~Sahidullah, J.~Patino, A.~Nautsch,
  X.~Liu, K.~A. Lee, T.~Kinnunen, N.~Evans \emph{et~al.}, ``Asvspoof 2021:
  accelerating progress in spoofed and deepfake speech detection,'' \emph{arXiv
  preprint arXiv:2109.00537}, 2021.

\bibitem{wavefake}
J.~Frank and L.~Sch{\"o}nherr, ``{WaveFake: A Data Set to Facilitate Audio
  Deepfake Detection},'' in \emph{Thirty-fifth Conference on Neural Information
  Processing Systems Datasets and Benchmarks Track}, 2021.

\bibitem{sgdr}
I.~Loshchilov and F.~Hutter, ``{SGDR:} stochastic gradient descent with warm
  restarts,'' in \emph{5th International Conference on Learning
  Representations, {ICLR} 2017, Toulon, France, April 24-26, 2017, Conference
  Track Proceedings}.\hskip 1em plus 0.5em minus 0.4em\relax OpenReview.net,
  2017.

\bibitem{lfcc_mfcc}
X.~Zhou, D.~Garcia-Romero, R.~Duraiswami, C.~Espy-Wilson, and S.~Shamma,
  ``Linear versus mel frequency cepstral coefficients for speaker
  recognition,'' in \emph{2011 IEEE workshop on automatic speech recognition \&
  understanding}.\hskip 1em plus 0.5em minus 0.4em\relax IEEE, 2011, pp.
  559--564.

\bibitem{pgdl2}
\BIBentryALTinterwordspacing
A.~Madry, A.~Makelov, L.~Schmidt, D.~Tsipras, and A.~Vladu, ``Towards deep
  learning models resistant to adversarial attacks,'' in \emph{International
  Conference on Learning Representations}, 2018. [Online]. Available:
  \url{https://openreview.net/forum?id=rJzIBfZAb}
\BIBentrySTDinterwordspacing

\bibitem{fab}
F.~Croce and M.~Hein, ``Minimally distorted adversarial examples with a fast
  adaptive boundary attack,'' in \emph{International Conference on Machine
  Learning}.\hskip 1em plus 0.5em minus 0.4em\relax PMLR, 2020, pp. 2196--2205.

\bibitem{Hendrycks_2021_ICCV}
D.~Hendrycks, S.~Basart, N.~Mu, S.~Kadavath, F.~Wang, E.~Dorundo, R.~Desai,
  T.~Zhu, S.~Parajuli, M.~Guo, D.~Song, J.~Steinhardt, and J.~Gilmer, ``The
  many faces of robustness: A critical analysis of out-of-distribution
  generalization,'' in \emph{Proceedings of the IEEE/CVF International
  Conference on Computer Vision (ICCV)}, October 2021, pp. 8340--8349.

\bibitem{NEURIPS2021_e19347e1}
Y.~Bai, J.~Mei, A.~L. Yuille, and C.~Xie, ``Are transformers more robust than
  cnns?'' in \emph{Advances in Neural Information Processing Systems},
  M.~Ranzato, A.~Beygelzimer, Y.~Dauphin, P.~Liang, and J.~W. Vaughan, Eds.,
  vol.~34.\hskip 1em plus 0.5em minus 0.4em\relax Curran Associates, Inc.,
  2021, pp. 26\,831--26\,843.

\bibitem{Liu_2022_CVPR}
Y.~Liu, Y.~Cheng, L.~Gao, X.~Liu, Q.~Zhang, and J.~Song, ``Practical evaluation
  of adversarial robustness via adaptive auto attack,'' in \emph{Proceedings of
  the IEEE/CVF Conference on Computer Vision and Pattern Recognition (CVPR)},
  June 2022, pp. 15\,105--15\,114.

\bibitem{mcd}
R.~Kubichek, ``Mel-cepstral distance measure for objective speech quality
  assessment,'' in \emph{Proceedings of IEEE Pacific Rim Conference on
  Communications Computers and Signal Processing}, vol.~1, 1993, pp. 125--128
  vol.1.

\bibitem{SzegedyZSBEGF13}
C.~Szegedy, W.~Zaremba, I.~Sutskever, J.~Bruna, D.~Erhan, I.~J. Goodfellow, and
  R.~Fergus, ``Intriguing properties of neural networks,'' in \emph{2nd
  International Conference on Learning Representations, {ICLR}}, 2014.

\bibitem{ijcai2021p591}
\BIBentryALTinterwordspacing
T.~Bai, J.~Luo, J.~Zhao, B.~Wen, and Q.~Wang, ``Recent advances in adversarial
  training for adversarial robustness,'' in \emph{Proceedings of the Thirtieth
  International Joint Conference on Artificial Intelligence, {IJCAI-21}}, 8
  2021, pp. 4312--4321. [Online]. Available:
  \url{https://doi.org/10.24963/ijcai.2021/591}
\BIBentrySTDinterwordspacing

\bibitem{tramer2018ensemble}
F.~Tramèr, A.~Kurakin, N.~Papernot, I.~Goodfellow, D.~Boneh, and P.~McDaniel,
  ``Ensemble adversarial training: Attacks and defenses,'' in
  \emph{International Conference on Learning Representations}, 2018.

\end{thebibliography}

\end{document}